\documentstyle[12pt]{article}
\renewcommand{\theequation}{\arabic{section}.\arabic{equation}}
\renewcommand{\baselinestretch}{1.5}

\begin{document}

\begin{flushright}
UT-Komaba 95-9
\end{flushright}

\begin{center} 
{\Large{\bf Finite Temperature Properties of \\
\vskip 0.3cm  
the Gauge Theory of Nonrelativistic Fermions  }}
\vskip 1.5cm

{\Large  Masaru Onoda$^{\dagger}$, 
Ikuo Ichinose$^{\dagger}$\footnote{e-mail 
 address: ikuo@hep1.c.u-tokyo.ac.jp}, and 
Tetsuo Matsui$^{\ast}$\footnote{e-mail
 address: matsui@phys.kindai.ac.jp}}  
\vskip 0.5cm
 
$^{\dagger}$Institute of Physics, University of Tokyo, 
Komaba, Tokyo, 153 Japan  \\
$^{\ast}$Department of Physics, Kinki University, 
Higashi-Osaka, 577 Japan

\end{center}

\begin{center} 
\begin{bf}
Abstract
\end{bf}
\end{center}
We study the finite temperature properties of the  gauge 
theory of nonrelativistic fermions introduced by Halperin, Lee, and Read.
This gauge theory is relevant to two interesting systems:
high-$T_c$ superconductors in the anomalous metallic phase 
and a two-dimensional electron 
system in a strong magnetic field
at the Landau filling factor $\nu=1/2$.
We calculate the self-energies of both gauge bosons  and  
fermions by the random-phase approximation, showing 
that the  dominant term at low energies is generated by
the gauge-fermion interaction.
The current-current correlation function is also calculated
by the ladder approximation. We confirm that the electric 
conductivity satisfies the Drude formula and obtain  
its temperature dependence, which is of a  non-Fermi-liquid.

\newpage

\section{Introduction}

In the last several years, it has been well recognized that gauge 
field theories play important roles in some interesting topics 
in condensed matter physics, like the fractional 
quantum Hall effect (FQHE) and the high-$T_c$ superconductivity.
One of the canonical low-energy models describing such electron 
systems is believed to be a U(1) gauge theory of nonrelativistic  
fermions, although its concrete form has not been identified yet.   

In the strongly-correlated electron systems like  the high-$T_c$ 
superconductors, it is expected that the  phenomenon of charge-spin 
separation (CSS) takes place
at low temperatures ($T$) \cite{Anderson}, that is, the charge 
and spin degrees of freedom of electrons behave independently.
Various experiments are explained consistently by assuming the CSS.
In the previous papers \cite{IMCSS}, two of the present authors
showed that the CSS can be explained very naturally by   
a confinement-deconfinement phase transition of strong-coupling
gauge theory. As demonstrated there, the CSS occurs at low 
 $T$ and the quasi-excitations there
are holons, spinons, and gauge bosons.

For the FQHE,  a Ginzburg-Landau (GL) theory has been 
proposed \cite{Girvin}, which explains various experimental results.
This GL theory is a Chern-Simons (CS) gauge theory coupled with a 
complex boson field; so-called bosonized electrons, and the FQH 
state is characterized as a condensation of the bosonized 
electrons. Motivated by the success of the GL theory as well as
Jain's idea of composite fermions  for the FQHE \cite{Jain}, 
Halperin, Lee, and Read \cite{HLR} studied  the system of electrons 
at the Landau filling factor $\nu = 1/2$ by introducing and analyzing 
a U(1) gauge theory of nonrelativistic fermions. This  theory contains 
a parameter $b$, which controls  the strength of gauge-field  fluctuations 
[see (\ref{model}),(\ref{VB}) in Sect.2]. 

Because of gauge invariance, the transverse component of gauge 
boson may survive as a massless mode, i.e., not shielded by vacuum 
polarization due to  fermions, and so  fermions interacting massless  
gauge bosons  may have non-Fermi-liquid behavior at low energies.
With this expectation, the effect of gauge field on the low-energy 
fermionic excitations has been studied by the random-phase 
approximation (RPA) \cite{HLR,Ioffe} and by the 
renormalization-group (RG) equation \cite{IMRG}.
The dynamics of  transverse gauge field is controlled by the Landau 
dissipative (damping) term in its propagator.
At $T=0$, the fermions exhibit marginal-Fermi-liquid
like behavior \cite{Varma} due to the coupling to this gauge field.
This phenomenon bears a close resemblance to the coherent 
soft-photon dressing of electrons in  quantum electrodynamics
 (QED). The latter is known to be crucial for resolving the 
problem of infrared singularities in QED.

Therefore, ample  systematic studies have been carried out so far
for this gauge theory in case of  $T=0$.
In this paper, we shall study its finite-temperature properties.
There have appeared some studies on similar topics: 
Lee and Nagaosa \cite{LN} calculated the conductivity in the uniform 
RVB mean-field theory
plus gauge field fluctuations of  the t-J model of high-$T_c$ 
superconductivity. 
This case corresponds to the special  value $b = 0$. 
Kim {\it et al.} \cite{Kim} calculated 
the current-current correlation functions at $T=0$  at the two-loop level, 
and get the conductivity at finite $T$ by assuming the Drude formula and  
certain scaling arguments. 
We shall compare our methods and results with theirs 
in some details. They are summarized in  
Sect.4.3 and in Sect. 5.

This paper is organized as follows.
In Sect.2, we introduce the gauge  model, which is, as announced,
relevant to the metallic phase of high-$T_c$ superconductors and 
the electron system at $\nu = 1/2$.
The RG studies of the model at $T=0$ \cite{IMRG} shows 
that there is a nontrivial infrared (IR) fixed point, whose 
location depends on the parameter $b$.
This fixed point describes a non-Fermi-liquid.
In Sect.3, we calculate the self-energies of both
 fermion and gauge-field propagators by RPA,
to show that the relevant term  at low energies appears through
the loop corrections.
In Sect.4, the current-current correlation function is calculated
by the ladder approximation (LA).
It is shown that the Drude formula of the conductivity is derived.
By using the Kubo formula, we obtain the $T$-dependence of  dc 
conductivity in the leading order of low $T$.
It exhibits non-Fermi-liquid behavior for $0 \leq b \leq 1$. 
In short, the resistivity for $b=1$ behaves as 
$\rho(T) \propto T^2 |\ln  T | $ . 
For $  b \neq  1$, we employ the  $\varepsilon$-expansion w.r.t. 
$ \varepsilon \equiv 1-b$.  For $ 0 \leq b < 1$, it behaves as   
$\rho(T) \propto T^{4/(3-b)} $. For $1 < b < 2$,  the Fermi-liquid behavior
is obtained, i.e., $\rho(T) \propto T^{2} $.
Special attentions are paid to the gauge invariance of the results.
Section 5 is devoted for conclusions. In Appendix, detailed
calculations of the current-current correlation function and 
the conductivity are presented.


\section{Model}

We shall consider a two-dimensional system of  nonrelativistic  
spinless fermion $\psi(x,\tau)$ interacting with a dynamical gauge  field 
$A_i(x,\tau)$ $(i=1,2)$.  In the imaginary-time formalism,  the action  
of the model at finite $T$ is given by 
\begin{eqnarray}
S&=&\int^{\beta}_0d\tau\int d^2 x\left\{\bar{\psi} (\partial_{\tau}-\mu)\psi
+\frac1{2m}\overline{(D_i\psi)} (D_i\psi)\right\} \nonumber\\
 & &+\frac12\int^{\beta}_0 d\tau \int d^2 x\; d^2 y\;   B(x,\tau)V_B (x-y)B(y,\tau), 
\label{model}
\end{eqnarray}
where $\beta \equiv (k_B T)^{-1}$. The covariant derivative $D_i$  
and the magnetic field $B$ are given by
\begin{equation}
D_i=\partial_i+igA_i,\quad B=\epsilon_{ij}\partial_i A_j.
\end{equation}
The fluctuations of $B$ are controlled by the "potential" function, 
\begin{eqnarray}
V_B(x-y)&=&v_B\int\underline{dq} \frac{e^{iq\cdot(x-y)}}{q^b},  \label{VB}  \\
\int \underline{dq}&=&{d^2q \over (2\pi)^2}, \nonumber \end{eqnarray}
where  the  parameter $b$ is assumed to be in the region   $0\leq b<2 $.
For the high-$T_c$ superconductivity, $b$ is chosen as  $b=0$ \cite{LN}, 
and for the electron system of the half-filled  Landau level,  $b$ is related to 
the  Coulombic-type repulsion between electrons \cite{HLR,IMRG}, 
but not yet fixed uniquely. We have also introduced the parameter $v_B$ 
for dimensional reason.

We shall take the Coulomb gauge $\partial_i A_i=0$.
The vector potential is then expressed in terms of $B$ as
\begin{equation}
A_i(x,\tau) =\epsilon_{ij}\int d^2 y\; \partial_j {\cal G}(x-y)B(y,\tau), 
\label{gaugef}
\end{equation}
\begin{equation}
{\cal G}(x)=\int \underline{dq}\frac{\exp(iq\cdot x)}{q^2}.
\end{equation}
We treat $B(x,\tau)$ as a fundamental dynamical field,  instead of $A_i(x,\tau)$ itself.
By substituting (\ref{gaugef}) into (\ref{model}),
\begin{eqnarray}
S&=&\int^{\beta}_0 d\tau\int d^2 x\left\{\bar{\psi} (\partial_{\tau}-\mu)\psi
+\frac1{2m}\partial_i\bar{\psi} \partial_i\psi\right\} \nonumber\\
 & &+\frac12\int^{\beta}_0 d\tau\int d^2 x \; d^2 y\;   
 B(x,\tau)V_B(x-y)B(y,\tau) \nonumber\\
 & &-\int^{\beta}_0 d\tau\int d^2 x \; d^2 y\;
   \epsilon_{ij}\partial_j{\cal G}(x-y)B(y,\tau)
\frac{ig}{2m}\bar{\psi}\stackrel{\leftrightarrow} {\partial_i}\psi(x,\tau) \nonumber\\
 & &+\int^{\beta}_0 d\tau\int d^2 x\; d^2 y\; d^2 z\;
   \epsilon_{ij}\partial_j{\cal G}(x-y)
    \epsilon_{ik}\partial_k{\cal G}(x-z) \nonumber\\
 & & \quad\times B(y,\tau)B(z,\tau)\frac{g^2}{2m}  \bar{\psi}\psi(x,\tau).
 \label{model2}
\end{eqnarray}
We perform Fourier transformation for $\psi (x,\tau)$  and $B(x,\tau)$, 
\begin{eqnarray}
\psi(x,\tau)&=&\frac1{\beta}\sum_{n}\int\underline{dk}\;
 \exp(ik\cdot x-i\omega_n\tau)\psi(k,\omega_n) \\
B(x,\tau)   &=&\frac1{\beta}\sum_{n}\int\underline{dq}\;
 \exp(iq\cdot x-i\epsilon_n\tau)qA(q,\epsilon_n),
\end{eqnarray}
where
\begin{equation}
\omega_n\equiv\frac{(2n+1)\pi}\beta,\quad \epsilon_n\equiv 
\frac{2n\pi}\beta, \quad n \in 
\mbox{{\bf Z}}.
\end{equation}
Then the action (\ref{model2}) becomes
\begin{eqnarray}
S&=&\frac1\beta\sum_n\int \underline{dk}\;\bar{\psi}(k,\omega_n)
 \left\{-i\omega_n+\frac{k^2}{2m}-\mu\right\}\psi(k,\omega_n)
    \nonumber\\
 & &+\frac12\frac1{\beta}\sum_n\int\underline{dq}\;   
 A(-q,-\epsilon_n)v_Bq^{2-b}A(q,\epsilon_n) \nonumber\\
 & &+\frac1{\beta^2}\sum_{n,l}\int\underline{dk}\;\underline{dq}\;  
\frac{ig}{m}\frac{k\times q}{q} A(q,\epsilon_l)  
\bar{\psi}(k+q,\omega_n+\epsilon_m)\psi(k,\omega_n) \nonumber\\
 & &+\frac1{\beta^3}\sum_{n,l,l'}\int \underline{dk}\; 
 \underline{dq}\;\underline{dq'}\;   \frac{g^2}{2m}\frac{(-q\cdot q')}{qq'} 
 A(q,\epsilon_l)A(q',\epsilon_{l'})\nonumber\\
 & &\quad\times  \bar{\psi}(k+q+q',\omega_n+\epsilon_l  
+\epsilon_{l'})\psi(k,\omega_n),
\label{model3}
\end{eqnarray}
where 
\begin{eqnarray}
k\times q\equiv \epsilon_{ij}k_i q_j =k q \sin \phi,\nonumber\\  
k\cdot q\equiv k_i q_i = k q \cos \phi.
\end{eqnarray}

In Sect.3, we shall study how the fermion and the gauge field  propagators 
are renormalized by the gauge-fermion interactions, 
$A\bar{\psi}\psi$ and $A A \bar{\psi}\psi$.


\section{Self-energies of fermions and gauge bosons}
\setcounter{equation}{0}
In this section we calculate the self-energies of the fermion and  
gauge-field propagators at finite $T$  by employing the RPA.
At $T=0$, it has been shown that the loop corrections generate the relevant 
terms at low energies \cite{HLR,IMRG}.

From (\ref{model3}), the gauge field propagator at the  tree level is given by
\begin{equation}
D_0^{-1}(q,\epsilon_l)=v_Bq^{2-b}.
\end{equation}
In the RPA, the diagrams in Fig.1 are summed up as a geometric  
series in order to obtain the corrected 
propagator  $D(q,\epsilon_l)$ of the gauge field:
\begin{eqnarray}
D^{-1}(q,\epsilon_l)&=&D_0^{-1}(q,\epsilon_l) +\Pi(q,\epsilon_l) \nonumber\\
                    &=&v_B q^{2-b}+\Pi(q,\epsilon_l).
\label{D}
\end{eqnarray}
By the straightforward calculation, we obtain
\begin{eqnarray}
\Pi(q,\epsilon_l)&=&\Pi_M+\widetilde{\Pi}(q,\epsilon_l)  \label{self0},  \\
\Pi_M&=&\frac{g^2}m\;\frac1\beta\sum_n \int\underline{dk}\;G_0(k,\omega_n) 
\nonumber\\
 &=&\frac{g^2}m\int\underline{dk}\;f_\beta(E(k))  \label{self1},\\
\widetilde{\Pi}(q,\epsilon_l)
&=&\left(\frac{g}{m}\right)^2\frac1\beta\sum_n \int\underline{dk}
\left(\frac{k\times q}{q}\right)^2
   G_0(k,\omega_n)G_0(k+q,\omega_n+\epsilon_l) \nonumber\\
&=&\left(\frac{g}{2\pi m}\right)^2\int^{\infty}_0  dk\;k^3
\int^{\pi}_{-\pi}d\phi\;\sin^2\phi\;
   \frac{f_\beta(E(k))-f_\beta(E(k+q))}{i\epsilon_l-\Delta E(q,k)},      
\label{self2}
\end{eqnarray}
where
\begin{eqnarray}
G^{-1}_0(k,\omega_n)&=&i\omega_n-E(k),\label{G_0}\\
f_\beta(E)&=&\frac1{e^{\beta E}+1},\\  
E(k)&=&\frac{k^2}{2m}-\mu,  \label{fermidis}
\end{eqnarray}
and 
\begin{eqnarray}
\Delta E(q,k)&=&E(k+q)-E(k)\nonumber\\ 
&=&\frac{kq}m\cos\phi+\frac{q^2}{2m}.
\end{eqnarray} 

In the later discussion, we shall assume  the conditions $\beta^{-1}\ll \mu$ 
and $q\ll k_F$.  These imply that we consider 
low-energy excitations near the Fermi surface.
In this case we have
\begin{equation}
\Pi_M\nonumber\\ =\frac{g^2}{2\pi\beta} \ln(1+e^{\beta\mu})\simeq 
\frac{g^2\mu}{2\pi},
\label{PiM}
\end{equation}
\begin{eqnarray}
\widetilde{\Pi}(q,\epsilon_l)
&=&-\left(\frac{g}{2\pi m}\right)^2\int^{\infty}_0 dk 
\int^{\pi}_{-\pi}d\phi  \frac{k^3\sin^2\phi}{i\epsilon_l -\Delta E(q,k)} \nonumber\\
& \times&\left[\Delta E(q,k)\frac{\partial  f_\beta}{\partial E}(E(k))  
+\frac{\{\Delta E(q,k)\}^2}2 \frac{\partial^2 f_\beta}{\partial E^2}(E(k))
+\cdots\right]
   \nonumber\\
&\simeq&-\Pi_M+\frac{g^2\mu}{\pi}
  \left\{\frac1{4\mu}\left(i\epsilon_l+\frac{q^2}{2m}\right) 
+\left(\frac{\epsilon_l}{2\pi i} -\frac{{\epsilon_l}^2}{4\pi\mu}\right)
I\left(\frac{q^2}{2m},\frac{k_F q}{m},\epsilon_l\right)\right\}, \nonumber\\
& &
\label{Pitilde}    
\end{eqnarray}
where 
\begin{eqnarray}
I(u,v,w)&=&\int^{\pi}_{-\pi}d\phi\frac{\sin^2\phi} {u+v\cos\phi-iw} \nonumber\\
 &=&\frac{i}{2v}\oint_{C_1}dz\frac{(z-z^{-1})^2}  
 {z^2+2\left(\frac{u-iw}{v}\right)z+1}.
\end{eqnarray}
For $0 < u \ll v$ , $I(u,v,w)$ is evaluated as follows:
\begin{equation}
I(u,v,w)\simeq\frac{2\pi i}{v^2}\left\{\mbox{sgn}(w) (v^2+w^2)^{\frac12}-iu\right\}
\left\{1-\frac{|w|}{(v^2+w^2)^{\frac12}}\right\}.
\label{Iuvw0}	
\end{equation}
From (\ref{self0}), (\ref{Pitilde}) and (\ref{Iuvw0}), we obtain
\begin{eqnarray}
\Pi(q,\epsilon_l)&\simeq&\frac{g^2\mu}{\pi}
 \left\{\frac1{4\mu}\left(i\epsilon_l+\frac{q^2}{2m}\right)
+\left(\frac{\epsilon_l}{2\pi i} -\frac{{\epsilon_l}^2}{4\pi\mu}\right)
    I\left(\frac{q^2}{2m},\frac{k_F q}{m},     \epsilon_l\right)\right\} \nonumber\\
&\simeq&\frac{g^2\mu}{\pi}\cdot \frac{|\epsilon_l|}{v_F q}\left[\left\{1+
\left( \frac{|\epsilon_l|}
{v_F q}\right)^2\right\}^\frac12
-\frac{|\epsilon_l|}{v_F q}\right],
\label{gaugerpa}
\end{eqnarray}
where  $v_F\equiv k_F/ m$. Therefore $\Pi(q, \epsilon_l)$ behaves as

(1) $\epsilon_l\ll v_F q$
\begin{equation}
\Pi(q,\epsilon_l)\simeq \frac{g^2\mu}{\pi} \cdot\frac{|\epsilon_l|}{v_F q},
\label{dissipative}
\end{equation}

(2) $\epsilon_l\gg v_F q$
\begin{equation}
\Pi(q,\epsilon_l)\simeq \frac{g^2\mu}{2\pi}.
\label{mass}
\end{equation}
Eq.(\ref{dissipative}) is nothing but the Landau damping factor,  
which plays an important role at $T=0$.
The above result shows that, at low $T$,  the term (\ref{dissipative}) is  
dominant, because $\epsilon_l\propto T$ and the summation over $l$ 
goes up to $l\sim \beta v_Fq$.
On the other hand, at high $T$, the effect of  the  dissipative 
term is less efficient.
For the high-$T_c$ superconductivity, the above remark is  important for 
the discussion on the confinement-deconfinement  phase transition (CDPT).
Actually, by using the hopping expansion \cite{IMCSS},  
it is shown that the CDPT occurs   in the t-J model  at a finite 
critical temperature, $T_{CD} > 0$.  This result is strongly related with  
the above remark on the dissipative term.
The CDPT in the present model is under study,  
and the results will be reported in future publications.

By using the gauge field propagator (\ref{gaugerpa}) obtained by the  RPA, 
we shall calculate the corrected fermion propagator $G(k,\omega_n)$.
The corresponding diagram is given in Fig.2, which gives rise to 
\begin{eqnarray}
G^{-1}(k,\omega_n)&=&G_0^{-1}(k,\omega_n) -\Sigma(k,\omega_n) 
\nonumber\\
    &=&i\omega_n-E(k)-\Sigma(k,\omega_n),
\label{fermionpro1}		  
\end{eqnarray}
\begin{eqnarray}
\Sigma(k,\omega_n)
&=&\left(\frac{g}{m}\right)^2\frac1\beta\sum_l\int\underline{dq}\;  
\left(\frac{k\times q}{q}\right)^2
   G_0(k+q,\omega_n+\epsilon_l)D(q,\epsilon_l) \nonumber\\
&=&-\left(\frac{gk}{2\pi m}\right)^2 \frac1\beta\sum_l\int^{k_F}_0 dq\;q\;
   \frac{I\left(\frac{k^2}{2m}+\frac{q^2}{2m}-\mu,\frac{kq}{m},   
\omega_n+\epsilon_l\right)}{v_B q^{2-b}+\Pi(q,\epsilon_l)},
\end{eqnarray}

\begin{equation}
\Sigma(k_F,\omega_n)
\simeq -i\cdot\frac{g^2v_F}{2\pi}\cdot\frac1\beta\sum_l \int^{k_F}_0 dq\;
   \frac{\mbox{sgn}(\omega_n+\epsilon_l) \left[\left\{1+\left(\frac{|\omega_n
+\epsilon_l|}{v_F q}\right)^2 \right\}^\frac12 -\frac{|\omega_n+\epsilon_l|}
{v_F q}\right]} {v_B q^{2-b} +\frac{g^2\mu}{\pi}\cdot\frac{|\epsilon_l|} 
{v_F q}\left[\left\{1+\left(\frac{|\epsilon_l|}{v_F q}\right)^2 \right\}^\frac12 
-\frac{|\epsilon_l|}{v_F q}\right]},
\label{Sigma2}
\end{equation}

We evaluate the $q$-integral in (\ref{Sigma2}) as follows.
Let us assume that the dominant region  satisfies $v_F q \ll 2\pi \beta^{-1}$.
This gives rise to the peak of the integrand to be $q \sim k_F$, which
contradicts the assumption for low $T$. On the other hand,
if  the dominant region is assumed to satisfy the opposite
inequality, $v_F q \gg 2\pi \beta^{-1}$, the peak of the integrand,
$q^{3-b}  \propto |\epsilon_l|$ (for $b < 2$) brings no incompatibility.
With this assumption, the integral is simplified as follows:
\begin{eqnarray}
\Sigma(k_F,\omega_n)
&\simeq& -i\cdot\frac{\mu}{\pi}\cdot\frac{\pi^2\alpha}{\beta\mu}
\sum_l\int^1_0 d\tilde{q}\;
   \frac{\mbox{sgn}(\Omega_n+{\cal E}_l)}
{\tilde{q}^{2-b}+\frac{|{\cal E}_l|}{\tilde{q}}} \nonumber\\
&\simeq& -i\cdot\frac{\mu}{\pi}\cdot\mbox{sgn}(\Omega_n)
\int^{|\Omega_n|}_{-|\Omega_n|}d{\cal E}\int^1_0 d\tilde{q}\;
   \frac{\tilde{q}}{\tilde{q}^{3-b}+|{\cal E}|} \nonumber\\
&=&-i\cdot\frac{2\mu}{\pi}\cdot\mbox{sgn}(\Omega_n)\int^1_0 d\tilde{q}\;\tilde{q}
   \ln\left(1+\frac{|\Omega_n|}{\tilde{q}^{3-b}}\right) \nonumber\\
&=&-i\cdot\frac{2\mu}{\pi}\cdot\mbox{sgn}(\Omega_n)|\Omega_n|^{\frac2{3-b}}
H_b(|\Omega_n|),
\label{qint}
\end{eqnarray}
where we have introduced the following 
dimensionless variables for  later convenience: 
\begin{eqnarray}
&&\alpha\equiv\frac{g^2 v_F}{2\pi^2 v_B{k_F}^{1-b}},
\quad \tilde{q}\equiv\frac{q}{k_F}, \nonumber\\
&&\Omega_n \equiv\frac{\pi\alpha}2\cdot\frac{\omega_n}\mu, \quad 
{\cal E}_l\equiv\frac{\pi\alpha}2\cdot\frac{\epsilon_l}\mu.
\end{eqnarray}
$H_b(c)$ in the final line  of (\ref{qint}) is given by
\begin{eqnarray}
H_b(c)
&\equiv&\frac1{3-b}\int^{\infty}_{c}dy\; y^{\frac{1-b}{3-b}-2}\ln(1+y)\nonumber\\
&\simeq&\frac1{1-b}\left(1-c^{\frac{1-b}{3-b}}\right) 
\qquad \mbox{for}\quad c\ll 1 \mbox{ and } b\sim 1.
\label{Hbc}
\end{eqnarray}
We get the second  line of  (\ref{qint})  by replacing the $l$-sum 
with an integral and using  the formula:
\begin{eqnarray}
& &\int^{\infty}_{-\infty} dx\; \mbox{sgn}(a+x) f(|x|) \nonumber\\
&=&\theta(a)\left\{\int^{\infty}_{-|a|}dx\; f(|x|)
-\int^{-|a|}_{-\infty}dx\; f(|x|)\right\}\nonumber\\
& &+\theta(-a)\left\{\int^{\infty}_{|a|}dx\;  f(|x|)
-\int^{|a|}_{-\infty}dx\; f(|x|)\right\} \nonumber\\
&=&\mbox{sgn}(a)\int^{|a|}_{-|a|}dx\; f(|x|).
\end{eqnarray}
Especially for $b=1$, we obtain
\begin{equation}
\Sigma(k_F,\omega_n)
\simeq -i\cdot\frac{\mu}{\pi}\cdot\Omega_n
\left\{\ln\left(1+\frac1{|\Omega_n|}\right)
+\frac1{|\Omega_n|}\ln\left(1+|\Omega_n|\right)\right\}.
\label{ologo} \end{equation}
By using the above results, we shall calculate the current-current  
correlation functions by the LA in the following section.

\section{Current-current correlation function and the conductivity}
\setcounter{equation}{0}
 
In the previous section, we obtained the corrected gauge and  the 
fermion propagators at finite $T$.
The loop corrections generate nontrivial relevant terms  at low energies. 
The low-energy behavior of the fermion propagator has a branch cut  
rather than a pole in the frequency, and this behavior has a close  resemblance 
to the 1D Luttinger liquid and the  over-screened Kondo effect. 
Therefore, one can expect that the gauge-fermion  interaction generates 
non-Fermi-liquid behavior also in  gauge-invariant correlation functions.
In these non-Fermi liquid systems, the $T$-dependence  of the resistivity 
$\rho$ behaves as $\rho (T) \propto T^{\Delta},  \; \Delta < 2$, which is different 
from that of the usual  Fermi liquid theory $\rho (T) \propto T^2$.
We expect similar properties for the present gauge system.

In this section, we shall calculate the current-current correlation  
function (CCCF) at finite $T$ by the LA.
At $T=0$, this correlation was calculated by Kim {\it et al.} \cite{Kim}  
at the two-loop order. They observed important cancellation of the leading 
singularities between the fermion self-energy  and the vertex correction,  
due to the gauge invariance. In the LA below, we shall also evaluate the  
diagrams  corresponding to their calculations,
 i.e., the fermion self-energy and  the vertex correction.   

\subsection{Schwinger-Dyson equation} 
The gauge-invariant electromagnetic current $J_i(x, \tau)$ is given by 
\begin{eqnarray}
J_i(x, \tau)&=&j_i(x, \tau) 
   -\frac{g}{m}\bar{\psi}\psi A_i(x, \tau)\nonumber\\
  j_i(x, \tau) &\equiv&\frac{i}{2m}\left\{\bar{\psi}\cdot\partial^i\psi(x, \tau)
  -\partial^i\bar{\psi}\cdot\psi(x, \tau)\right\}.
\end{eqnarray}  
The effect of the second (contact) term in $J_i(x, \tau)$ on the 
conductivity is less dominant at low $T$. This can be seen in a straightforward
manner from the calculations by Kim {\it et al.} \cite{Kim}.
Therefore we consider  the CCCF for $j_i(q, \epsilon_l)$ that  is given by 
\begin{eqnarray}
&&\langle j_i(q,\epsilon_l) j_j(q',\epsilon_{l'})\rangle \nonumber\\
&=&\frac1{\beta^2}\sum_{n,n'}\int\underline{dk}\;\underline{dk'}\;
 \frac{2k_i+q_i}{2m}\cdot\frac{2{k'}_j+{q'}_j}{2m} \nonumber\\
& &\times\langle \bar{\psi}(k,\omega_n)\psi(k+q,\omega_n+\epsilon_l)
    \bar{\psi}(k',\omega_{n'})\psi(k'+q',\omega_{n'}+\epsilon_{l'})      
    \rangle \nonumber\\
&=&-\bar{\delta}(q+q')\;\beta\delta_{l,-l'}\Pi_{ij}(q,\epsilon_l). 
\end{eqnarray}
It satisfies the "Schwinger-Dyson (SD)"  equation which is graphically 
depicted  in Fig.3.   To calculate the conductivity, 
we shall use the  Kubo formula. In that calculation, only the limit 
$q,\; q'  \rightarrow 0$ of the above CCCF is needed.
Therefore, we focus on the CCCF at  zero-momenta below.

In order to solve the SD equation in the LA, it is useful to  start with the 
following expression for the polarization tensor $\Pi_{ij}(0, \epsilon_l)$:
\begin{equation}
\Pi_{ij}(0,\epsilon_l)
=\frac1{\beta^2}\sum_{n,n'}\int\underline{dk}\;\underline{dk'}\; 
\frac{k_i}{m}\cdot\frac{{k'}_j}{m}\; Y(k,\omega_n;k',\omega_{n'}).
\end{equation}
In term of the above function $Y(k,\omega_n;k', \omega'_n;\epsilon_m)$, 
the SD equation is rewritten as 
\begin{eqnarray}
& &Y(k,\omega_n;k',\omega_{n'};\epsilon_l)\nonumber\\
&=&R(k,\omega_n;\epsilon_l)\bar{\delta}(k-k') \beta\delta_{n,n'} \nonumber\\
& &+ R(k,\omega_n;\epsilon_l)\frac1\beta\sum_{n''} \int\underline{dk''}
   \Biggl[\left(\frac{g}{m}\right)^2 \left(\frac{k\times k''}{|k-k''|}\right)^2
  D(k''-k,\omega_{n''}-\omega_n)\nonumber\\
& &+\bar{\delta}(k-k'')\;\beta\delta_{n,n''}
\left\{\Sigma(k,\omega_n) {G_0}^{-1}(k'',\omega_{n''}+\epsilon_l)
+ \Sigma(k'',\omega_{n''} +\epsilon_l){G_0}^{-1}(k'',\omega_{n''})\right\}\Biggr] \nonumber\\
& &\times Y(k'',\omega_{n''};k',\omega_{n'};\epsilon_l),
\label{SD}
\end{eqnarray}
where 
\begin{equation}
R(k,\omega_n;\epsilon_l)\equiv  G_0(k,\omega_n)\;G_0(k,\omega_n+\epsilon_l),
\label{R}
\end{equation}
and $G_0(k,\omega_n)$ is the fermion propagator at the tree level  defined 
in (\ref{G_0}). To reduce the SD equation to more tractable form,
we consider the following integral of the function $Y$:
\begin{equation}
C_j(k,\omega_n;\epsilon_l)=\frac1{\beta}\sum_{n'}\int \underline{dk'}\;
{k'}_{j} Y(k,\omega_n;k',\omega_{n'};\epsilon_l).
\label{C}
\end{equation}
In terms of this $C_j(k,\omega_n;\epsilon_l)$, the SD equation (\ref{SD}) becomes
\begin{eqnarray}
& &C_j(k,\omega_n;\epsilon_l) \nonumber\\
&=&k_jR(k,\omega_n;\epsilon_l) 
+\frac{\Delta\Sigma(k,\omega_n;\epsilon_l)}{i\epsilon_l} C_j(k,\omega_n;\epsilon_l) 
\nonumber\\
& &+\left(\frac{g}{m}\right)^2\frac1\beta\sum_{n''} \int\underline{dk''}   
\left(\frac{k\times k''}{|k-k''|}\right)^2  D(k''-k,\omega_{n''}-\omega_n) \nonumber\\
& &\qquad\times\left\{R(k,\omega_n;\epsilon_l) C_j(k'',\omega_{n''};\epsilon_l)   
-R(k'',\omega_{n''};\epsilon_l)  C_j(k,\omega_n;\epsilon_l)\right\},
\label{SD1}
\end{eqnarray}
where we write
\begin{equation}
\Delta\Sigma(k,\omega_n;\epsilon_l)\equiv  R^{-1}(k,\omega_n;\epsilon_l)
\left\{\Sigma(k,\omega_n) {G_0}^2(k,\omega_n) 
-\Sigma(k,\omega_n+\epsilon_l) {G_0}^2(k,\omega_n+\epsilon_l)\right\}.
\label{DeltaSig}
\end{equation}
In the above we used the relation:
\begin{eqnarray}
& &R(k,\omega_n;\epsilon_l)\frac1{\beta}\sum_{n''} \int\underline{dk''}\;
\bar{\delta}(k''-k)\;\beta\delta_{n,n''}
    C_j(k'',\omega_{n''};\epsilon_l)\nonumber\\
& &\times  \left\{\Sigma(k,\omega_n){G_0}^{-1}(k'',\omega_{n''} +\epsilon_l)   
+\Sigma(k'',\omega_{n''}+\epsilon_l) {G_0}^{-1}(k'',\omega_{n''})\right\} \nonumber\\
&=&\frac{R^{-1}(k,\omega_n;\epsilon_l)}{i\epsilon_l} 
\left\{\Sigma(k,\omega_n){G_0}^2(k,\omega_n)
   -\Sigma(k,\omega_n+\epsilon_l){G_0}^2(k,\omega_n+\epsilon_l)\right\}
C_j(k,\omega_n;\epsilon_l) \nonumber\\
& &-\frac1{\beta}\sum_{n''}\int\underline{dk''}\;\left(\frac{g}{m} \right)^2
\left(\frac{k\times k''}{|k-k''|}\right)^2
 D(k''-k,\omega_{n''}-\omega_n)R(k'',\omega_{n''};\epsilon_l)
     C_j(k,\omega_n;\epsilon_l).\nonumber
\end{eqnarray}

Now, let us make the following ansatz for  $C_j(k,\omega_n;\epsilon_l)$ 
to solve the SD equation (\ref{SD1}),
\begin{equation}
C_j(k,\omega_n;\epsilon_l)=k_j R(k,\omega_n;\epsilon_l) 
\Psi(k,\omega_n;\epsilon_l),
\label{Psi}
\end{equation} where $\Psi(k,\omega_n;\epsilon_l)$ is the unknown function to be determined. 
This form is natural owing to the rotational symmetry. 
Then, $\Psi(k,\omega_n;\epsilon_l)$ must satisfy  
\begin{eqnarray}
&&\Psi(k,\omega_n;\epsilon_l)\nonumber\\
&=& 1+\frac{\Delta\Sigma(k,\omega_n;\epsilon_l)}{i\epsilon_l} 
\Psi(k,\omega_n;\epsilon_l) \nonumber\\
& & +\left(\frac{g}{m}\right)^2\frac1\beta\sum_{n''} \int\underline{dk''}   
\left(\frac{k\times k''}{|k-k''|}\right)^2  D(k''-k,\omega_{n''}-\omega_n)  \nonumber\\
&& \quad \times R(k'',\omega_{n''};\epsilon_l) \left\{\frac{k\cdot k''}{k^2} 
\Psi(k'',\omega_{n''};\epsilon_l)-\Psi(k,\omega_n;\epsilon_l)\right\},
\label{SDPsi}
\end{eqnarray}
where we have used the fact that the following integral $\Gamma_j(k,\omega_n;\epsilon_l)$  
is proportional to $k_j$,
\begin{eqnarray}
\Gamma_j(k,\omega_n;\epsilon_l)
&=&\left(\frac{g}{m}\right)^2\frac1\beta\sum_{n''}\int\underline{dk''}
   \left(\frac{k\times k''}{|k-k''|}\right)^2 \nonumber\\  
&& \times  D(k''-k,\omega_{n''}-\omega_n) \; 
k''_{j} R(k'',\omega_{n''};\epsilon_l)  \Psi(k'',\omega_{n''};\epsilon_l) \nonumber\\
&=& k_j\;\frac{k\cdot \Gamma(k,\omega_n;\epsilon_l)}{k^2}. \nonumber  
\end{eqnarray}
[In the last line, the definition of $\Gamma_j(k,\omega_n;\epsilon_l)$ is used.] 
At this stage, the CCCF is expressed as
\begin{equation}
\Pi_{ij}(0,\epsilon_l)
=\frac1{\beta}\sum_n\int\underline{dk}\;
\frac{k_i}{m}\cdot \frac{k_j}{m}\;R(k,\omega_n;\epsilon_l)\Psi(k,\omega_n;\epsilon_l).
\label{CCCFC}
\end{equation} 
Hence, by solving (\ref{SDPsi}) for $\Psi(k,\omega_n;\epsilon_l)$, 
we get a solution for CCCF.

To solve (\ref{SDPsi}) we first note that the $k''$ integral in (\ref{SDPsi})
  is dominated by the region 
$k''\sim k_F$ due to the appearance of  $R(k'',\omega_n;\epsilon_l)$ 
as long as $\Psi(k,\omega_n;\epsilon_l)$ is a smooth function of $k$. 
Furthermore, we make an assumption that the $n$-dependence 
of  $\Psi(k_F,\omega_n;\epsilon_l)$ is weak. 
One shall see that this crucial assumption is satisfied in the final solution.
So this is a self-consistent solution.
With these simplifications, the solution of (\ref{SDPsi}) is easily obtained as 
\begin{equation}
\Psi(k_F,\omega_n;\epsilon_l)
=\frac{i\epsilon_l}{i\epsilon_l-i\epsilon_l \Gamma_{GI}(k_F,\omega_n;\epsilon_l)  
-\Delta\Sigma(k_F,\omega_n;\epsilon_l)},
\label{sol}
\end{equation}
where 
\begin{eqnarray}
\Gamma_{GI}(k,\omega_n;\epsilon_l)
&\equiv&\left(\frac{g}{m}\right)^2\frac1\beta\sum_{n''} \int\underline{dk''}   
\left\{\frac{k\times (k''-k)}{|k''-k|}\right\}^2  \frac{k\cdot (k''-k)}{k^2}  \nonumber\\
& &\qquad\times D(k''-k,\omega_{n''}-\omega_n)  R(k'',\omega_{n''};\epsilon_l). 
\label{GGI}
\end{eqnarray}
\subsection{The case  of $b = 1$}
Below we consider the region of low $T$ to get the concrete results. 
we shall discuss the case $b=1$ first, because this case allows us to 
extract the leading nontrivial term of  $\Psi(k_F,\omega_n;\epsilon_l)$ at low $T$.
Let us evaluate $\Gamma_{GI}(k_F,\omega_n;\epsilon_l)$ as follows:
we present basic steps of calculations and explain the approximations involved.
The reader who are interested in more details can find them in Appendix.
First, by using the 
polar coordinate, and setting $|k|, |k^{''}| = k_F$ in $D(k''-k,\omega_{n''}-\omega_n)$, 
we get 
\begin{eqnarray}
\Gamma_{GI}(k_F,\omega_n;\epsilon_l)
&\simeq&-\frac{g^2\mu}{(2\pi)^2}\cdot\frac1\beta\sum_{n''}
    \int^{\pi}_{-\pi}d\phi\;\sin^2\phi  \;D\!\left(k_F\sqrt{2(1-\cos\phi)}, 
\omega_{n''}-\omega_n\right) \nonumber\\
& &\qquad\times \int^{\infty}_{-\infty}dE\;\frac1{i\omega_{n''} -E}\cdot
\frac1{i\omega_{n''} +i\epsilon_l -E} \nonumber\\
&=&-\frac{g^2\mu}{2\pi}\cdot\frac1{|\epsilon_l|}\cdot\frac1\beta{\sum_{n''}}'
    \int^{\pi}_{-\pi}d\phi\;\sin^2\phi \nonumber\\
& &\qquad\times  D\!\left(k_F\sqrt{2(1-\cos\phi)},\omega_{n''}-\omega_n\right),
\end{eqnarray}
where the contour integral over $E$ restricts the $n^{''}$-sum; ${\sum}'$ denotes  
the summation over $n^{''}$ satisfying 
 $\mbox{sgn}(\omega_{n''}) =-\mbox{sgn}(\epsilon_l),\; |\omega_{n''}|<|\epsilon_l|$. 

By repeating the similar argument as in the $k''$ integral in (\ref{Sigma2}) above,
we find that one should use the damping term (\ref{dissipative}) for $D(q, \epsilon_l)$ rather than 
the mass term (\ref{mass}). Then we get
\begin{eqnarray}
\Gamma_{GI}(k_F,\omega_n;\epsilon_l)
&\simeq&-\frac{g^2\mu}{2\pi}\cdot\frac1{|\epsilon_l|}\cdot\frac1\beta{\sum_{n''}}'
    \int^{\pi}_{-\pi}d\phi\;\sin^2\phi \nonumber\\
& &\qquad\times\frac{ k_F\sqrt{2(1-\cos\phi)}}
{2v_B k_F^2 (1-\cos\phi)+\frac{g^2\mu}{\pi v_F}|\omega_{n''}-\omega_n|} \nonumber\\ 
&=&-\frac{\alpha}4\cdot\frac1{|{\cal E}_l|}\cdot\frac{\pi^2\alpha}{\beta\mu}
{\sum_{n''}}' Q_1 (|\Omega_{n''}-\Omega_n|),
\end{eqnarray}
where
\begin{eqnarray}
Q_1 (c)&\equiv&\int^{\pi}_{-\pi}d\phi\;
       \frac{\sin^2\phi\;\sqrt{2(1-\cos\phi)}}{2(1-\cos\phi)+c} 
      \nonumber\\
    &=&\int^2_0 dx\;\frac{x^3\sqrt{4-x^2}}{x^{2}+c}.
\end{eqnarray}
 To perform the $\phi$-integral or $x$-integral, $x \equiv \{2(1-\cos\phi)\}^{1/2}$, 
we note that $Q_1 (0)$ can be exactly evaluated. For small $c$,
a scaling argument gives rise to the following leading behavior:
\begin{equation}
Q_1(c) \simeq \frac{8}{3} +B \; c  \ln c,
\end{equation}
where $B$ is some constant. 
Then we have
\begin{equation}
\Gamma_{GI}(k_F,\omega_n;\epsilon_l)
\simeq -\frac{\alpha}4\cdot\frac1{|{\cal E}_l|} \cdot\frac{\pi^2\alpha}{\beta\mu}
{\sum_{n''}}'\left\{\frac83-B|\Omega_{n''}-\Omega_n|\ln| 
\Omega_{n''}-\Omega_n|^{-1}\right\}.
\label{GV1}
\end{equation}  
Since the $n''$-sum is restricted and $T$ is small, we ignore the weak $n''$ and $n$ 
 dependence in the log factor above by replacing it by 
$\ln\left\{\pi^2 \alpha/(\beta \mu)\right\}^{-1}$. 
Then the $n''$-sum is carried out explicitly to reach the final result, 
\begin{eqnarray}
\Gamma_{GI}(k_F,\omega_n;\epsilon_l)
&\simeq&-\frac{\alpha}4\left[ \frac{8}{3}- B\;\mbox{sgn}({\cal E}_l)\;
\left(\frac{\pi^2\alpha}{\beta\mu}\right)^2 \ln 
\left(\frac{\pi^2\alpha}{\beta\mu}\right)^{-1}
\left\{\frac{l(l+1)}2+(l+1)n+n^2 \right\}\right] .\nonumber\\
\label{GGIfinal}
\end{eqnarray}  
From (\ref{sol}) and (\ref{GGIfinal}),  we get
\begin{eqnarray}
\Psi(k_F,\omega_n;\epsilon_l)
&\simeq&\frac{i{\cal E}_l}{i\tilde{C}_1(\beta){\cal E}_l 
+i\gamma_b(k_F,\omega_n;\epsilon_l)},
\label{sol1}   \\
\gamma_b(k_F,\omega_n;\epsilon_l)
&\simeq&-\frac{\alpha}4 B\;\mbox{sgn}({\cal E}_l)\;
\left(\frac{\pi^2\alpha}{\beta\mu}\right)^2 
\ln \left(\frac{\pi^2\alpha}{\beta\mu}\right)^{-1}
\left\{\frac{l(l+1)}2+(l+1)n+n^2 \right\} ,\nonumber\\
\tilde{C}_1(\beta) &\simeq& 1+\frac{\alpha}{2}\left\{\frac{4}{3}
 +\ln \left(\frac{\pi^2\alpha}{\beta \mu}\right)^{-1}\right\}.
 \label{tau}
\end{eqnarray} The second term of the coefficient $\tilde{C}_1(\beta)$ 
comes from $\Delta \Sigma(k_F,\omega_n;\epsilon_l)$ of (\ref{DeltaSig}), 
which reflect the behavior of 
the fermion self-energy (\ref{ologo}),  
\begin{equation}
\Sigma(k,\omega_n) \sim \omega_n \ln |\omega_n|^{-1}. 
\label{Sigma1}
\end{equation}
It is straightforward to see  that
$\Delta \Sigma(k_F,\omega_n;\epsilon_l)$ is proportional to $\epsilon_l$ for $k= k_F$.

The solution (\ref{sol1}) has certainly a weak  $n$-dependence  
in the relevant region of $|\Omega_n| < |{\cal E}_l|$, 
$\mbox{sgn}(\Omega_n) =-\mbox{sgn}({\cal E}_l)$ for $\Pi_{ij}(0, \epsilon_l)$ 
at low $T$ in a self-consistent manner as we assumed.
In Fig.4 we plotted $\Omega_n$-dependence of $\Gamma_{GI}(k_F,\omega_n;\epsilon_l)$, 
which supports this assumption. 
By inserting (\ref{sol1}) into (\ref{CCCFC}), we  obtain  the following result for the CCCF:
\begin{equation}
\Pi_{ij}(0,\epsilon_l)
\sim \delta_{ij}\;\frac{\mu}{2\pi}\cdot
\frac{i\epsilon_l}{i\tilde{C}(\beta)\epsilon_l +i\tau^{-1}(\beta)}
\qquad \mbox{for $\epsilon_l > 0$},
\label{Drude}
\end{equation}
where
\begin{equation}
\tau^{-1}(\beta)\simeq\frac{\mu}{6\pi} B \left(\frac{\pi^2\alpha}{\beta\mu}\right)^2
\ln\left(\frac{\pi^2\alpha}{\beta\mu}\right)^{-1}
\propto T^2 |\ln T|.
\label{tauB}
\end{equation}
  The electric conductivity $\sigma_{ij}$ is obtained by the analytic  continuation, 
\begin{equation}
\epsilon_l \rightarrow  -i\epsilon +\delta,
\label{analyticcont}
\end{equation}
where $\delta$ is an infinitesimal positive constant.
Then we observe that Eq.(\ref{Drude}) is nothing but 
the Drude formula  of the electric conductivity.  
Explicitly, the dc conductivity is given as 
\begin{eqnarray}
\mbox{Re}\,\sigma_{ij}&=&\lim_{\epsilon \rightarrow 0  \nonumber  }  
\frac{e^2}{-i\epsilon} \;\Pi_{ij}(0,-i\epsilon+\delta) \nonumber  \\
&\simeq& \delta_{ij}\frac{e^2\rho}{m}\;\tau(\beta).
\label{conductivity}
\end{eqnarray}
In the conventional Fermi-liquid theory, the resistivity  behaves as 
$\rho(T) \propto T^2$.
In contrast,  in the present case, $\rho(T) \propto T^2 |\ln T|$.
This difference supports that the present system with $b=1$  
is a  "marginal" Fermi liquid.
\subsection{ The case of $0 \le b < 1$}
Next, let us consider the case $b < 1 $  at low $T$. To obtain a concrete result, 
we employ the idea of $\varepsilon$  expansion in the critical phenomena. 
Here we use $\varepsilon \equiv 1-b$ as  a small expansion parameter to 
expand various quantities around  the known ones at $b=1$. 
The detailed calculations are  collected in Appendix. 
The expression of $\Gamma_{GI}(k_F,\omega_n;\epsilon_l)$ is obtained as follows:
\begin{equation}
\Gamma_{GI}(k_F,\omega_n;\epsilon_l)
\simeq -\frac{\alpha}4\cdot\frac1{|{\cal E}_l|}\cdot\frac{\pi^2\alpha}{\beta\mu}
{\sum_{n''}}'\left\{A_b-
B_b\left(|\Omega_{n''}-\Omega_n|\right)\,|\Omega_{n''}-\Omega_n|\right\},
\end{equation}
where 
\begin{equation}
B_b(c)\simeq\frac{B}{1-b}\left\{c^{-\frac{2(1-b)}{3-b}}-1\right\}
\qquad \mbox{for $c\ll 1$ and $b\sim 1$},
\label{Bbc}
\end{equation}
and $A_b$ is a constant which depends on $b$.
We plot $\Gamma_{GI}(k_F,\omega_n;\epsilon_l)$ as a function of $\omega_n$ in Fig.4. 
This exhibits that it has a weak
$n$-dependence, as we assumed to obtain the solution.
It should be also remarked that $\Delta \Sigma(k_F,\omega_n;\epsilon_l)$  of 
(\ref{DeltaSig}) getting very large for $\Omega_n \sim 0$ or $-{\cal E}_l$,
 and  the region in which $\Delta \Sigma(k_F,\omega_n;\epsilon_l)$
is small scales linearly with respect to $|{\cal E}_l|$. 
This behavior is confirmed numerically as Fig.5 shows.
As in the case of $b=1$, the momentum integral in (\ref{CCCFC})  
can be carried out and obtain the restriction on the summation over $n$  
in the interval $|\Omega_n| < |{\cal E}_l|$, 
$\mbox{sgn}(\Omega_n) =-\mbox{sgn}({\cal E}_l)$.
With these remarks, we again obtain the Drude formula for 
conductivity (\ref{Drude}) and its explicit $T$-dependence. 
The final results of conductivity  at low $T$ in the $(1-b)$-expansion is given by   
\begin{eqnarray}
\mbox{Re}\,\sigma_{ij}&\simeq& \delta_{ij} \frac{e^2\rho}{m} \;\tau(\beta), 
\nonumber\\
\tau^{-1}(\beta)&\equiv&\frac{\mu}{6\pi}\left(\frac{\pi^2\alpha}{\beta\mu}\right)^2
                  B_b\!\left(\frac{\pi^2\alpha}{\beta\mu}\right)
                 \propto T^{\frac{4}{3-b}}.
\label{tau2}
\end{eqnarray}

From this result, we conclude that the conductivity behaves as 
that of a non-Fermi-liquid for general values of $0 \leq b < 1$.
As the model of high-$T_c$ superconductivity, the parameter $b=0$ is chosen, 
and so we get
$\rho(T)\propto T^{4/3}$ in the present calculation.
This result coincides with the calculation of the relaxation time  obtained by 
Lee and Nagaosa \cite{LN}.
In their calculation, however, it is not clear whether   
the gauge invariance is respected or not.
On the other hand, in our calculation, 
the final expressions of  the conductivity, (\ref{tauB}) and 
(\ref{tau2}), come  from the quantity $\Gamma_{GI}(k_F,\omega_n;\epsilon_l)$ 
in $\Psi(k_F,\omega_n;\epsilon_l)$. 
This $\Gamma_{GI}(k_F,\omega_n;\epsilon_l)$ is a gauge-invariant
combination of the fermion-self-energy and the vertex correction. 
Strictly speaking, the quantity $\Psi(k_F,\omega_n;\epsilon_l)$ itself should be gauge  invariant 
because the current itself is gauge invariant (up to
the irrelevant contact terms). However, due to the LA, 
the  other term $\Delta \Sigma(k_F,\omega_n;\epsilon_l)$ 
appears in the denominator of $\Psi(k_F,\omega_n;\epsilon_l)$, 
which is not gauge invariant.   However, after taking
 the limit $\epsilon \rightarrow 0$, it does not contribute  to the conductivity, 
 a physical quantity, as we have shown.
As explained,  the gauge-invariant CCCF's were also examined up to
 the two-loop order at $T=0$ by Kim {\it et al.} \cite{Kim}.
By assuming the Drude formula and some scaling rules,  
they obtained the $T$-dependence of conductivity 
for general values of $0 \leq b <1$.
Their result agrees with ours. The lesson one can learn from the present 
calculation is that the LA with the one-loop kernel in SD equation naturally 
leads to the Drude formula at low $T$ without any further assumptions.
We believe the $(1-b)$-expansion employed here  works well down to $b=0$. 
\subsection{ The case of $1 < b < 2$} To close this section,  we cite the result also for $1 < b < 2$ 
which is calculated in Appendix. 
The resistivity behaves as $\rho(T) \propto T^2$, i.e., 
just as the behavior of a Fermi-liquid.  This sharp asymmetry 
under $\varepsilon \leftrightarrow -\varepsilon$ comes from the behavior of 
the function $B_b$ of (\ref{Bbc}). 

\section{Conclusion}
\setcounter{equation}{0}

In this paper, we studied the finite-temperature properties of  
the gauge theory of nonrelativistic fermions.
The self-energies of the gauge field and the fermion were calculated  by the RPA, 
and it was found that relevant terms at low energies  
appear from the loop corrections.
We calculated also the CCCF by the LA.
We verified how the Drude formula is satisfied, and obtained 
the  $T$-dependence of the dc conductivity.
It coincides with that obtained by Lee and Nagaosa \cite{LN} for the 
special case of the high-$T_c$
superconductivity, i.e. $b=0$, and also the two-loop calculations by 
Kim {\it et al.} \cite{Kim} for the CCCF which assumed the 
Drude formula and some scalings. 
We stress that, as in the $T=0$ case of Kim {\it et al.},  
the cancellation takes place in the CCCF  between the singular part in the 
self-energy  of fermion and that of vertex correction  of fermion-gauge coupling.
This cancellation is reflected in the result 
$\rho(T)\propto T^{\frac{4}{3-b}} 
(0 \leq b < 1)$ at low $T$. This is less dominant than the naive result  
$\rho(T)\propto T^{\frac{2}{3-b}}$ \cite{Gan} that one may obtain by considering 
only the fermion propagator and ignoring the vertex corrections.
It is interesting to apply the present method of the LA together with 
the $(1-b)$-expansion to a  system of bosons interacting with the gauge field. 
Such a system is relevant also for the high-$T_c$ superconductivity.

Very recently, Nayak and Wilczek \cite{NWRG} made 
a RG study to discuss the finite-temperature properties of metals 
like one-dimensional Luttinger liquid, but not  
of gauge theories. We are studying the RG equations at finite $T$ of  
the present gauge theory as an extension of the previous analysis 
at $T=0$ \cite{IMRG}. The results will be  reported in a future 
publication \cite{TOIM}. 

Finally we comment on the recent experiment \cite{exp} 
of the  mobility at $\nu = 1/2$. They fit their data in the form of 
$C_1 +C_2 T^2 +C_3 |\ln  T | $,  where the last term takes care of  the effect of 
gauge-field scattering in a dirty metal with impurities calculated by
Halperin, Lee and Read \cite{HLR} .
The fitting  looks for us not so definitive to reject out all the other possibilities.
Our calculations in the present paper is for a clean metal with no impurities.
The similar calculations for a disordered system with impurities are under study.
It is an interesting subject to compare such experiments with our result  
in a systematic way. Such comparison will  certainly shed 
some  light on the  low-energy effective gauge theory of electrons at $\nu = 1/2$, 
e.g., by selecting out the best value of $b$.  

\newpage
\setcounter{equation}{0}

\renewcommand{\theequation}{A.\arabic{equation}}
\renewcommand{\baselinestretch}{1.5}

\section*{Appendix}

In this Appendix, we shall present detailed calculations of $\Gamma_{GI}(k_F,\omega_n;\epsilon_l)$ 
and  $\Pi_{ij}(0, \epsilon_l)$ for general values of $b$ assuming that $b \sim 1$ and $T$ is small. 
First, let us start with $\Gamma_{GI}(k,\omega_n;\epsilon_l)$ of (\ref{GGI}),
\begin{eqnarray}
\Gamma_{GI}(k,\omega_n;\epsilon_l)
&\equiv&\left(\frac{g}{m}\right)^2\frac1\beta\sum_{n''}\int\underline{dk''}
   \left\{\frac{k\times (k''-k)}{|k''-k|}\right\}^2 \frac{k\cdot (k''-k)}{k^2}  \nonumber\\
& &\qquad\times   D(k''-k,\omega_{n''}-\omega_n) R(k'',\omega_{n''};\epsilon_l),
\end{eqnarray}
where $D(q,\epsilon_l)$ is given by (\ref{D}) and (\ref{gaugerpa}).
By using the polar coordinate for the momentum integral,
\begin{eqnarray}
\Gamma_{GI}(k_F,\omega_n;\epsilon_l)
&\simeq&-\frac{g^2\mu}{(2\pi)^2}\cdot\frac1\beta\sum_{n''}
    \int^{\pi}_{-\pi}d\phi\;\sin^2\phi \; D\!\left(k_F\sqrt{2(1-\cos\phi)},
 \omega_{n''}-\omega_n\right) \nonumber\\
& &\qquad\times \int^{\infty}_{-\infty}dE\;\frac1{i\omega_{n''} -E}
\cdot\frac1{i\omega_{n''} +i\epsilon_l -E} \nonumber\\
&=&-\frac{g^2\mu}{2\pi}\cdot\frac1{|\epsilon_l|}\cdot\frac1\beta{\sum_{n''}}'
    \int^{\pi}_{-\pi}d\phi\;\sin^2\phi \nonumber\\
& &\qquad\times  D\!\left(k_F\sqrt{2(1-\cos\phi)},\omega_{n''}-\omega_n\right).
\end{eqnarray}
In the dominant region of the above integral, 
$v_F k_F\sqrt{2(1-\cos\phi)}\gg 2\pi / \beta$, and so
\begin{eqnarray}
\Gamma_{GI}(k_F,\omega_n;\epsilon_l)
&\simeq&-\frac{g^2\mu}{2\pi}\cdot\frac1{|\epsilon_l|}\cdot\frac1\beta{\sum_{n''}}'
    \int^{\pi}_{-\pi}d\phi\;\sin^2\phi \nonumber\\
& &\qquad\times\frac{ k_F\sqrt{2(1-\cos\phi)}}
{v_B\left\{k_F\sqrt{2(1-\cos\phi)}\right\}^{3-b}+\frac{g^2\mu}{\pi v_F}
|\omega_{n''}-\omega_n|} \nonumber\\
&=&-\frac{\alpha}{4}\cdot\frac1{|\epsilon_l|}\cdot\frac{2\pi}\beta{\sum_{n''}}'
    \int^{\pi}_{-\pi}d\phi\;\sin^2\phi \nonumber\\
& &\qquad\times\frac{\sqrt{2(1-\cos\phi)}}
{\left\{\sqrt{2(1-\cos\phi)}\right\}^{3-b}+\frac{\pi\alpha }2
\cdot\frac{|\omega_{n''}-\omega_n|}{\mu}} \nonumber\\
&=&-\frac{\alpha}4\cdot\frac1{|{\cal E}_l|}\cdot\frac{\pi^2\alpha}{\beta\mu}
{\sum_{n''}}' Q_b(|\Omega_{n''}-\Omega_n|),
\end{eqnarray}
where
\begin{equation}
\alpha\equiv\frac{g^2v_F}{2\pi^2 v_B {k_F}^{1-b}},
\quad\Omega_n\equiv \frac{\pi\alpha}2\cdot\frac{\omega_n}{\mu},
 \quad {\cal E}_l\equiv \frac{\pi\alpha}2\cdot\frac{\epsilon_l}{\mu},
\end{equation}
and
\begin{eqnarray}
Q_b(c)&\equiv&\int^{\pi}_{-\pi}d\phi\;
       \frac{\sin^2\phi\;\sqrt{2(1-\cos\phi)}}{\left\{\sqrt{2(1-\cos\phi)}\right\}^{3-b}+c} 
      \nonumber\\
    &=&\int^2_0 dx\;\frac{x^3\sqrt{4-x^2}}{x^{3-b}+c}.
\end{eqnarray}

When $c \ll 1$ and $b\sim 1$, we can calculate $Q_b$ approximately.
First, we shall consider the following integral.
\begin{equation}
B_b(c)\equiv\frac{Q_b(0)-Q_b(c)}{c}=\int^2_0 dx\;\frac{x^b\sqrt{4-x^2}}{x^{3-b}+c},
\end{equation}
The maximum of the above integrand is at $x\sim c^{\frac1{3-b}}$, and
the value of the integrand at $x=1$ is very small in comparison with 
the value at the maximum. So the naive evaluation will gives
$B_b(c)\propto c^{-\frac{2(1-b)}{3-b}}$. On the other hand, in the case of 
$b=1$, the singularity of $c \to 0$ appears as $\ln c^{-1}$. Therefore, we
can evaluate the above integral more precisely as 
$B_b(c)\propto (1-b)^{-1}\left\{c^{-\frac{2(1-b)}{3-b}}-1\right\}$.
In this way, we  get 
\begin{equation}
Q_b(c)= A_b-B_b (c)\,c,
\end{equation}
where
\begin{equation}
A_b\equiv\int^2_0 dx\;x^b\sqrt{4-x^2},
\qquad B_b (c)\simeq
\frac{B}{1-b}\left\{c^{-\frac{2(1-b)}{3-b}}-1\right\},
\label{Bb}
\end{equation}
and $B$ is some numerical constant.

Hereafter we assume $\epsilon_l>0$ without loss of generality.
To obtain an explicit expression for the conductivity, we have to assume that
the parameter $b$ is close to $1$, i.e.,  $b=1-\varepsilon$ and $|\varepsilon|$ is 
infinitesimally small.
After getting the result by assuming that $\varepsilon$ is small, we put,
for example,  $\varepsilon=1$
for $b=0$ in a similar spirit to the usual $\varepsilon$-expansion. 
When $b=1-\varepsilon$, we can get
\begin{eqnarray}
\Gamma_{GI}(k_F,\omega_n;\epsilon_l)
&\simeq&-\frac{\alpha}4\cdot\frac1{|{\cal E}_l|}\cdot\frac{\pi^2\alpha}{\beta\mu}
{\sum_{n''}}'\left\{A_b-
B_b\left(|\Omega_{n''}-\Omega_n|\right)|\Omega_{n''}-\Omega_n|\right\} \nonumber\\
&\simeq&\frac{\alpha}4\left[-A_b
+\frac1{|{\cal E}_l|}\;B_b\!\left(\frac{\pi^2\alpha}{\beta\mu}\right)
\frac{\pi^2\alpha}{\beta\mu}
\left\{\sum^{|n+1|}_{l'=1}{\cal E}_{l'}
+\sum^{|l+n|}_{l'=1}{\cal E}_{l'}\right\}\right] \nonumber\\
&=&\frac{\alpha}4\left[-A_b
+\frac1{|{\cal E}_l|}\;B_b\!\left(\frac{\pi^2\alpha}{\beta\mu}\right)
\left(\frac{\pi^2\alpha}{\beta\mu}\right)^2
\left\{\sum^{|n+1|}_{l'=1}l'+\sum^{|l+n|}_{l'=1}l'\right\}\right] \nonumber\\
&=&\frac{\alpha}4\left[-A_b
+\cdot\frac1{|{\cal E}_l|}\;B_b\!\left(\frac{\pi^2\alpha}{\beta\mu}\right)
\left(\frac{\pi^2\alpha}{\beta\mu}\right)^2
\left\{\frac{l(l+1)}2+(l+1)n+n^2\right\}\right]
\nonumber\\
&=&\frac{\alpha}4\left[-A_b
+\frac12\cdot\frac1{|{\cal E}_l|}\;B_b\!\left(\frac{\pi^2\alpha}{\beta\mu}\right)
\left\{|\Omega_n|^2+|{\cal E}_l+\Omega_n|^2-2(\Omega_0)^2\right\}\right].
\nonumber\\
\end{eqnarray}
When we translated the first line of the above equation to the second line,
$n^{\varepsilon}\sim 1$ was used.

If the $n$-dependence of $\Psi(k_F,\omega_n;\epsilon_l)$ is weak,  we have
\begin{equation}
\Psi(k_F,\omega_n;\epsilon_l)
\simeq\frac{i{\cal E}_l}{iC_b{\cal E}_l+i\varsigma_b (k_F,\omega_n;\epsilon_l)
       +i\gamma_b(k_F,\omega_n;\epsilon_l)},
\end{equation}
where
\begin{equation}
C_b\equiv 1+\frac{\alpha}4 A_b \quad (A_0=\pi,A_1=\frac83),
\end{equation}
and
\begin{eqnarray}
\varsigma_b (k_F,\omega_n;\epsilon_l)
&\equiv &i\frac{\pi\alpha}{2\mu}\Delta\Sigma(k_F,\omega_n;\epsilon_l) \nonumber\\
     &\simeq&\alpha\left(\frac{\pi^2\alpha}{\beta\mu}\right)^{-\frac{1-b}{3-b}}
         H_b\!\left(\frac{\pi^2\alpha}{\beta\mu}\right)
         \Omega_n\left({\cal E}_l+\Omega_n\right)\left\{{\Omega_n}^{-1}
         -\left({\cal E}_l+{\Omega_n}\right)^{-1}\right\} \nonumber\\
      &=&\alpha\left(\frac{\pi^2\alpha}{\beta\mu}\right)^{-\frac{1-b}{3-b}}
           H_b\!\left(\frac{\pi^2\alpha}{\beta\mu}\right){\cal E}_l .
\end{eqnarray}	  
Here $H_b(c)$ is defined in (\ref{Hbc}) and estimated as
\begin{eqnarray}
H_b(c)&\simeq&\frac1{1-b}\left(1-c^{\frac{1-b}{3-b}}\right) 
        \qquad \mbox{for}\quad c\ll 1 \mbox{ and } b\sim 1,
\end{eqnarray}	
and 
\begin{eqnarray}
\gamma_b(k_F,\omega_n;\epsilon_l)
      &\equiv&-{\cal E}_l\left\{\Gamma_{GI}(k_F,\omega_n;\epsilon_l)
          +\frac{\alpha}{4}A_b\right\}\nonumber\\
      &\simeq&-\frac{\alpha}8\;B_b\!\left(\frac{\pi^2\alpha}{\beta\mu}\right)\;
      \mbox{sgn}({\cal E}_l)
      \left\{|\Omega_n|^2+|{\cal E}_l+\Omega_n|^2-2{\Omega_0}^2\right\}\nonumber\\
      &=&-\frac{\alpha}4\;B_b\!\left(\frac{\pi^2\alpha}{\beta\mu}\right)\;
      \mbox{sgn}({\cal E}_l)
      \left\{\frac{l(l+1)}2+(l+1)n+n^2\right\}.
\end{eqnarray}

Using the above results, $\Pi_{ij}(0, \epsilon_l)$ is given as follows:
\begin{eqnarray}
&&\Pi_{ij}(0,\epsilon_l) \nonumber \\
&\simeq&\delta_{ij}\frac{\mu}{2\pi}\cdot\frac1{|{\cal E}_l|}
\cdot\frac{\pi^2\alpha}{\beta\mu}{\sum_n}'\Psi(k_F,\omega_n;\epsilon_l)
\nonumber \\
&=&\delta_{ij}\frac{\mu}{2\pi}\cdot\frac{\pi^2\alpha}{\beta\mu}{\sum_n}'
     \frac{i\;\mbox{sgn}({\cal E}_l)}
{i\tilde{C}_b(\beta){\cal E}_l+i\gamma_b(k_F,\omega_n;\epsilon_l)}
\nonumber\\
&=&\delta_{ij}\frac{\mu}{2\pi}\cdot\frac{\pi^2\alpha}{\beta\mu}{\sum_n}'
     \frac1{\tilde{C}_b(\beta)|{\cal E}_l|}\left\{1-\frac1{\tilde{C}_b(\beta){\cal E}_l}
    \gamma_b(k_F,\omega_n;\epsilon_l)+\cdots\right\}
\nonumber\\
&\simeq&\delta_{ij}\frac{\mu}{2\pi}\cdot
     \frac1{\tilde{C}_b(\beta)|{\cal E}_l|}\left\{|{\cal E}_l|+\frac1{\tilde{C}_b(\beta)}
   \cdot\frac{\alpha}{12}B_b\!\left(\frac{\pi^2\alpha}{\beta\mu}\right)
\left({\cal E}_l+2\Omega_0\right)\left({\cal E}_l-2\Omega_0\right)+\cdots\right\}
\nonumber\\
&\simeq&\delta_{ij}\frac{\mu}{2\pi}\cdot
     \frac{|{\cal E}_l|}{\tilde{C}_b(\beta)|{\cal E}_l|
   -\frac{\alpha}{12}B_b\!\left(\frac{\pi^2\alpha}{\beta\mu}\right)
\left({\cal E}_l+2\Omega_0\right)\left({\cal E}_l-2\Omega_0\right) } \nonumber\\
&\simeq&\delta_{ij}\frac{\mu}{2\pi}\cdot
     \frac{i{\cal E}_l}{i\tilde{C}_b(\beta)\;{\cal E}_l
   +i\frac{\alpha}{12}B_b\!\left(\frac{\pi^2\alpha}{\beta\mu}\right)\mbox{sgn}({\cal E}_l)
    \;(i{\cal E}_l)^2+i\frac{\pi\alpha}{2\mu}\mbox{sgn}({\cal E}_l)\;\tau^{-1}(\beta)},
\end{eqnarray}
where
\begin{equation}
\tilde{C}_b(\beta)\equiv C_b
+ \alpha  \left(\frac{\pi^2\alpha}{\beta\mu}\right)^{-\frac{1-b}{3-b}}
H_b\!\left(\frac{\pi^2\alpha}{\beta\mu}\right), 
\end{equation}
\begin{equation}
\tau^{-1}(\beta)\equiv\frac{\mu}{6\pi}
\left(\frac{\pi^2\alpha}{\beta\mu}\right)^2 B_b\!\left(\frac{\pi^2\alpha}{\beta\mu}\right),
\label{Apptau}
\end{equation}
and we have used the formula
\begin{equation}
{\sum_n}'\left\{\frac{l(l+1)}2+(l+1)n+n^2 \right\}=\frac{\mbox{sgn}(l)}3 l(l+1)(l-1).
\end{equation}
Eq. (\ref{Apptau}) is the result given in the text. 
Using (\ref{Bb}), we can get concrete expressions for general $b\sim 1$
as follows:

(i) $0 \le b < 1$
\begin{equation}
\tau^{-1}(\beta) \simeq \frac{\mu}{6\pi}\cdot\frac{B}{1-b}
\left(\frac{\pi^2\alpha}{\beta\mu}\right)^2
\left\{\left(\frac{\pi^2\alpha}{\beta\mu}\right)^{-\frac{2(1-b)}{3-b}}-1\right\},
\end{equation}

(ii) $b=1$
\begin{equation}
\tau^{-1}(\beta) \simeq \frac{\mu}{6\pi}B
\left(\frac{\pi^2\alpha}{\beta\mu}\right)^2
\ln\left(\frac{\pi^2\alpha}{\beta\mu}\right)^{-1},
\end{equation}

(iii) $1 < b < 2$
\begin{equation}
\tau^{-1}(\beta)  \simeq \frac{\mu}{6\pi}\cdot\frac{B}{b-1}
\left(\frac{\pi^2\alpha}{\beta\mu}\right)^2 
\left\{1-\left(\frac{\pi^2\alpha}{\beta\mu}\right)^{\frac{2(b-1)}{3-b}}\right\}.
\end{equation}


\newpage
{\bf Figure Captions}
\begin{eqnarray}
&&\mbox{Figure 1: Graphs which contribute to the dressed gauge  
propagator in the RPA.}\nonumber \\
&&\mbox{Figure 2: A graph which contributes to the dressed  
fermion propagator.} \nonumber \\
&&\hspace{1.8cm}\mbox{The wavy line describes the dressed  
gauge boson propagator.} \nonumber  \\
&&\mbox{Figure 3: Graphical representation of the Schwinger-Dyson equation for the CCCFs.} \nonumber  \\
&&\mbox{Figure 4: Plots of  $\Gamma_{GI}$ as functions of $\Omega_n$.  
They have weak $\Omega_n$ dependence.}  \nonumber \\
&&\mbox{Figure 5: Plots of  $\Delta \Sigma$ as  functions of $\Omega_n$.  
They have peaks around $\Omega_n  \simeq 0 , -{\cal E}_l $.}  \nonumber
\end{eqnarray}

\end{document}